\documentclass[%
 reprint,
 amsmath,amssymb,
 aps,
]{revtex4-1}

\usepackage{graphicx}
\usepackage{dcolumn}
\usepackage{bm}
\usepackage[usenames, dvipsnames]{color}

\usepackage{cancel}

\begin{document}

\preprint{APS/123-QED}

\title{M\"{o}bius Quantum Walk}

\author{Majid Moradi}
\email{majidofficial@gmail.com}


\author{Mostafa Annabestani}
\email{annabestani@shahroodut.ac.ir}
\affiliation{Physics Department, Shahrood University of Technology.}


\date{\today}

\begin{abstract}
By adding an extra Hilbert space to Hadamard Quantum Walk on Cycles (QWC), we presented a new type of QWCs called M\"{o}bius Quantum Walk (MQW). The new space configuration enables the particle to rotate around the axis of movement. We defined factor $\alpha$ as the M\"{o}bius factor which is number of rotations per cycle. So by $\alpha=0$ we have normal QWC, while $\alpha \neq 0$ defines new type of QWC (namely M\"{o}bius Quantum Walk). Specially $\alpha = \frac{1}{2}$ defines a structure similar to M\"{o}bius strip. We analytically investigated this new type of QW and found that by tuning the parameter $\alpha$ we can reach uniform distribution for any number of nodes, while it is impossible for QWC. The effects of $\alpha$ on limiting distribution have been investigated and an explicit formula for non-uniform cases has been derived as well.

\end{abstract}

\pacs{Valid PACS appear here}
\maketitle

\section{\label{sec:level1}Introduction}
Introduced by Aharonov et al.\cite{Aharonov2001}, quantum walk (QW) on graphs is quantum counterpart of classical random walk (CRW) on graphs. While in CRW the particle moves with a certain probability, thanks to superposition in quantum mechanics, in QW a particle can move in all directions simultaneously \cite{Ambainis2001}.

In fact CRW is a dissipative model which its dispersion (variance of probability distribution) goes like $t$ (time)\cite{Feller1968}, while QW can be considered as a tight-binding model \cite{Nizama2012}, so its dispersion goes like $t^2$. The quadratic behavior of QW's variance is a direct consequence of unitary evolution of coherent QW, but in practice, the isolation of QW system from its environment is impossible. It is shown that the environmental effects and noises make the evolution to be non-unitary and decoherency happens \cite{Kendon2003,Nizama2012,Esposito2005}, So the variance of QW ($t^2$) may transits to the classical ones ($t$) \cite{Brun2003b,Annabestani2016}.

 Generally there are two types of QWs: Continuous-time QW \cite{Farhi1998} and discrete-time QW \cite{Watrous2001}.

Different types of QWs have been studied such as one-dimensional quantum walk \cite{Ambainis2001}, two-dimensional quantum walk \cite{Mackay2002} , quantum walk on graphs \cite{Aharonov2001} and hypercubes \cite{Moore2002} and a variety of parameters have been studied, such as hitting time \cite{Kempe2005}, mixing time \cite{Ambainis2001}, entanglement \cite{Abal2006}, decoherency \cite{Annabestani2010} , etc.

Aharonov \textit{et al.} \cite{Aharonov2001} designed a new model of QW known as Quantum Walk on Cycles (QWC), in which the discrete nodes can be supposed to be distributed on the circumference of a circle. They studied mixing time and limiting distribution (LD) of QWC and proved that for odd number of nodes the limiting distribution is uniform, while for even number of nodes it is not uniform. An explicit formula for non-uniform limiting distribution of QWC with even number of nodes has been driven by \cite{Bednarska2003, Portugal2013}. 
The study of mixing time as an important aspect of quantum walk is still interesting and new parameters such as transient temperature and its connection with mixing time has been introduced \cite{Diaz2016}.

We have modified QWC and add rotation ability to quantum walker and defined M\"{o}bius quantum walk in such a way that, while the particle walks along the cycle, it can rotate around the movement direction. We defined the parameter $\alpha$ to define number of rotations per cycle and investigated its effects on the parameters of QW. The explicit formula for limiting distribution has been derived.

This paper is organized as follows. Sec. \ref{sec:level2} provides a short introduction of QWC and limiting distribution of QWC. We also highlight two situations in QWC causing uniform and non-uniform distributions. In Sec. \ref{sec:level3} we introduce our model of QW namely MQW (Mobius Quantum Walk) and defined its evolution operator and solve its eigenprobelm. Finally in Sec. \ref{sec:level4} we find explicit formula for limiting distribution in degenerate (thus non-uniform) cases and discuss the role of M\"{o}bius factor in suppressing degeneracy and making the limiting distribution uniform.

\section{\label{sec:level2}Background}
Quantum walk (QW) and Classical random walk (CRW) are very similar, since there is a coin flip followed by a shift in position space \cite{Chandrashekar2008}, but quantum properties like superposition and interference \cite{Chandrashekar2009} cause the quantum walk to behave completely different.
  In one-dimensional discrete-time QW there are two Hilbert spaces known as coin $\mathcal{H_C}$ spanned by vectors $\left| s \right\rangle ,s = 0,1 $ and position space $\mathcal{H_{P}}$ spanned by vectors $\left| j \right\rangle $\cite{Ambainis2001}, so the state of walker is given by
\begin{equation}\label{ODQWState}
\left| {\psi _{t}} \right\rangle  = \sum\limits_{s = 0}^1 {\sum\limits_{j = 0}^{} {{\Psi _{s,j,t}} \left| {s,j} \right\rangle } }.
\end{equation}

One step of walk, $U=S ( I \otimes U_{C} )$, consists of two operators, coin operator $U_{c}$ which is a unitary $2 \times 2$ operator and makes superposition in coin space and shift operator $S$ which moves the walker according to the state of the coin
\begin{equation}\label{ODQWShift}
S = \sum\limits_{s = 0}^1 {\sum\limits_{j =  - \infty }^{ + \infty } {\left| {s,j + {{\left( { - 1} \right)}^s}} \right\rangle \left\langle {s,j} \right|} }.
\end{equation}

Different types of $U_{C}$ and $S$ introduce different types of QWs which have completely different properties. For example Ambainis et al. \cite{Ambainis2001} used Hadamard coin operator
\begin{equation}\label{Hadamard}
H = \frac{1}{{\sqrt 2 }}\left[ {\begin{array}{*{20}{c}}
1&1\\
1&{ - 1}
\end{array}} \right]
\end{equation}
and introduced 1-dimensional Hadamard walk and showed that it spreads linearly with number of steps, while for CRW it is $O\left( {\sqrt t } \right)$ \cite{Ambainis2001}. On the other hand changing the shift operator can also introduce new types of QWs. For example defining $S$ on a circle with finite number of nodes defines quantum walk on cycles (QWC)
\begin{equation}\label{QWCShiftOperator}
S = \sum\limits_{s = 0}^1 {\sum\limits_{j = 0}^{N - 1} {\left| {s,\left( {j + {{\left( { - 1} \right)}^s}} \right)modN} \right\rangle \left\langle {s,j} \right|} }.
\end{equation}
In which $N$ is number of the nodes. \\

Although it is shown that the probability distribution
\begin{equation}\label{QWCProbDist}
{p_t}\left( v \right) = \sum\limits_{s = 0}^1 {\left| {\left\langle {{s,v}}
 \mathrel{\left | {\vphantom {{s,v} {{\psi _t}}}}
 \right. \kern-\nulldelimiterspace}
 {{{\psi _t}}} \right\rangle } \right|^{2}}
\end{equation}
for QWC does not converge, but the limiting distribution $\pi (v)$ does \cite{Aharonov2001}, where
\begin{equation}\label{QWCLD}
\pi \left( v \right) = \mathop {\lim }\limits_{T \to \infty } {\bar p_T}\left( v \right)
\end{equation}
in which
\begin{equation}\label{QWCPBar}
{\bar p_T}\left( v \right)=\frac{1}{T}\sum\limits_{t = 1}^T {{p_t}\left( v \right)}.
\end{equation}

The limiting distribution for odd number of nodes is uniform, despite the form of the initial state and equals $\frac{1}{N}$, while for even number of nodes the limiting distribution is non-uniform and its explicit formula is \cite{Portugal2013}
\begin{equation}\label{EvNQWCLD}
\pi \left( v \right) = \frac{1}{N} + \frac{{{{\left( { - 1} \right)}^v}}}{{2{N^2}}}\sum\limits_{k = 0}^{N - 1} {\frac{{\cos (\frac{{4\pi k}}{N}v) - \cos (\frac{{4\pi k}}{N}\left( {v + 1} \right))}}{{1 + {{\cos }^2}\left( {\frac{{2\pi k}}{N}} \right)}}}.
\end{equation}

\begin{figure}
\centering
\includegraphics[width=90mm]{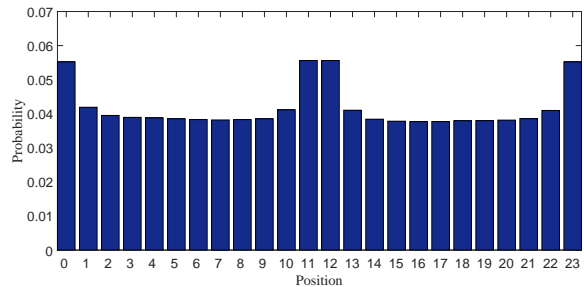}\hfill
\caption{\label{fig24NQWCLD} LD for Hadamard QWC with 24 nodes and initial state $| \psi _{0} \rangle = | 0 \rangle | 0 \rangle$}
\end{figure}

An example of limiting distribution of non-uniform cases has been plotted in Fig.~\ref{fig24NQWCLD}.

\section{\label{sec:level3}M\"{o}bius Quantum Walk}
In this section we modified QWC in such a way that the walker be able to rotate around the axis of movement while walking on the cycle. Seeking this goal, we added an extra Hilbert space $\mathcal{H}_{r}$ spanned by $ \left\lbrace  \left| r \right\rangle\mid r = 0,1 \right\rbrace $ to the Hilbert space of the QWC, so the state of the walker can be written as
\begin{equation}\label{MQWState}
\left| {{\psi _t}} \right\rangle  = \sum\limits_{s = 0}^1 {\sum\limits_{r = 0}^1 {\sum\limits_{j = 0}^{N - 1} {{\Psi _{s,r,j,t}}\left| {s,r,j} \right\rangle } } }
\end{equation}

We defined conditional rotation operator as
\begin{equation}\label{MQWRCond}
{R_{\theta}}\left| s \right\rangle \left| r \right\rangle  = {e^{ - \frac{i}{\hbar }\left( {{\sigma _z} \otimes \theta {S_x}} \right)}}\left| s \right\rangle \left| r \right\rangle,
\end{equation}
in which $\theta$ is the amount of rotation per step, $S_x=\frac{\hbar}{2}\sigma_x$ is spin operator along axis $x$; $\sigma_x$ and $\sigma_z$ are Pauli matrices.
We can rewrite \eqref{MQWRCond} as 
\begin{equation}\label{RthetaonKets}
{R_{\theta}}\left| s \right\rangle \left| r \right\rangle  = \left| s \right\rangle R\left(s\right)\left| r \right\rangle,
\end{equation}
with
\begin{equation}
R\left(s\right)=\cos\left( \frac{\theta}{2} \right){I_R} - i{{\left( { - 1} \right)}^s}\sin \left( {\frac{\theta}{2}} \right){\sigma _x}, 
\end{equation}
in which $I_{R}$ is identity operator in rotation space.

Conditional rotation operator $R_{\theta}$\cancel{,} rotates $\left| r \right\rangle$ along the axis $x$ according to the coin state $\left| s \right\rangle$. In other words, when the walker moves forward, $\left| r \right\rangle$ rotates counter-clockwise and when it moves backward, $\left| r \right\rangle$ rotates clockwise. So this structure can be interpreted as a twisted path like M\"{o}bius strip. That is why we call this model M\"{o}bius quantum walk.

In fact, an example of a M\"{o}bius strip can be considered as a paper strip, giving it a half-twist, and then joining the ends of the strip to form a loop. If a walker, walks along the length of this strip, it can visit both sides of the strip without crossing the edges. In fact, M\"{o}bius strip is a surface (defined by a normal vector to the surface) with only one side and only one boundary\cite{MobiusBook2005}. In our model, $\left|r\right\rangle$ plays the role of the normal vector to the surface of strip, so any node on the circumference of a circle not only has a specific index for its location but also a normal vector for the surface which it is located on.

By these modification, one step of MQW can be defined as
\begin{equation}\label{MQWU}
U=S \left( H \otimes I_{R} \otimes I_{P} \right),
\end{equation}
where act of $S$ is
\begin{equation}\label{MQWShift}
S|s,r,j\rangle =|s\rangle R(s) |r\rangle | ( j+(-1)^s) \, mod \, N \rangle
\end{equation}
and $H$ is Hadamard operator. By using Fourier transform
\begin{equation}\label{MQWVKappa}
| {{\kappa _k}} \rangle  = \frac{1}{{\sqrt N }}\sum\limits_{n = 0}^{N - 1} {{e^{\frac{{2i\pi kn}}{N}}} | n \rangle } 
\end{equation}
we can change the basis of representation and simplify our calculations. Therefore matrix elements of $U$ in k-space are
\begin{equation}\label{MQWUK}
\begin{array}{l}
\langle s,r,{\kappa _k}|U\left| {s',r',{\kappa _{k'}}} \right\rangle  =\\
\underbrace {\left( {{e^{\frac{{ - 2i\pi k}}{N}}}{H_{0,s'}}R(0) + {e^{\frac{{2i\pi k}}{N}}}{H_{1,s'}}R(1)} \right)}_M{\delta _{k,k'}}.
\end{array}
\end{equation}

In fact $M$ is the evolution operator of the system in k-space. Therefore after $t$ steps of walking we need to know $M^{t}$. We can use spectral decomposition and write $M$ as
\begin{equation}\label{MQWCUKSpectral}
M = \sum\limits_{s = 0}^1 {\sum\limits_{r = 0}^1 {\sum\limits_{k = 0}^{N - 1} {{e^{i{\lambda _{s,r,k}}}}\left| {{\lambda _{s,r,k}}} \right\rangle \left\langle {{\lambda _{s,r,k}}} \right|} } }
\end{equation}
After some calculations, one can find the eigenvalues as
\begin{equation}\label{EigVal}
\begin{array}{*{20}{c}}
{\begin{array}{*{20}{c}}
{{\lambda _{0,0,k}} =  - {e^{i\delta }}}\\
{{\lambda _{0,1,k}} = {e^{ - i\delta }}}
\end{array}}&{,\,\,\,\,\,\sin (\delta ) = \frac{1}{{\sqrt 2 }}sin\left( {\frac{{2\pi k}}{N} - \frac{\theta }{2}} \right)}\\
{\begin{array}{*{20}{c}}
{{\lambda _{1,0,k}} =  - {e^{i\sigma }}}\\
{{\lambda _{1,1,k}} = {e^{ - i\sigma }}}
\end{array}}&{,\,\,\,\,\,\sin (\sigma ) = \frac{1}{{\sqrt 2 }}sin\left( {\frac{{2\pi k}}{N} + \frac{\theta }{2}} \right)}
\end{array}
\end{equation}
and the corresponding eigenstates (see Appendix.\eqref{MQWEigStaApp}).

\section{\label{sec:level4}MQW Limiting Distribution}
If we assume initial state of the walker in k-space as
\begin{equation}\label{MQWIntStaK}
| {{{\tilde \psi }_0}} \rangle  = \sum\limits_{s = 0}^1 {\sum\limits_{r = 0}^1 {\sum\limits_{k = 0}^{N - 1} {{C_{s,r,k}}\left| {{\lambda _{s,r,k}}} \right\rangle } } },
\end{equation}
after $t$ steps we will have
\begin{equation}\label{MQWTSta}
| {{{\tilde \psi }_t}} \rangle  = M^{t} | {{{\tilde \psi }_0}} \rangle  = \sum\limits_{s = 0}^1 {\sum\limits_{r = 0}^1 {\sum\limits_{k = 0}^{N - 1} {{C_{s,r,k}}{e^{i{\lambda _{s,r,k}}t}}\left| {{\lambda _{s,r,k}}} \right\rangle } } }.
\end{equation}

The probability distribution on QWCs does not converge, but the limiting distribution \eqref{QWCPBar} as asymptotic time average of probability distribution in each node does \cite{Aharonov2001}. By using \eqref{QWCProbDist}, \eqref{QWCPBar} and \eqref{MQWTSta}
\begin{equation} \label{MQWLD1}
\begin{array}{l}
{\bar p_T}(v) = \sum\limits_{b,s,s' = 0}^1 {\sum\limits_{\rho ,r,r' = 0}^1 {\sum\limits_{k,k'}^{N - 1} {{C_{s,r,k}}C_{s',r',k'}^{*}\langle b,\rho ,v|{\lambda _{s,r,k}}\rangle } } }
\\
\,\,\,\,\,\,\,\,\,\,\,\,\,\,\,\,\,\, \langle {\lambda _{s',r',k'}}|b,\rho ,v\rangle  \times \frac{1}{T}\sum\limits_{t = 0}^{T - 1} {{e^{i \left( {{\lambda _{s,r,k}} - {\lambda _{s',r',k'}}} \right)t}}}
\end{array}
\end{equation}
In this equation only the last term is time-dependent. So for large T, by some algebra one can show that \cite{Aharonov2001}
\begin{equation}\label{MQWLD2}
\small{
\mathop {\lim }\limits_{T \to \infty } \frac{1}{T}\sum\limits_{t = 0}^{T - 1} {\left( {{e^{i \left( {{\lambda _{s,r,k}} - {\lambda _{s',r',k'}}} \right)t}}} \right) = \left\{ {\begin{array}{*{20}{c}}
0&{}&{\left( {{\lambda _{s,r,k}} \ne {\lambda _{s',r',k'}}} \right)}\\
{}&{}&{}\\
1&{}&{\left( {{\lambda _{s,r,k}} = {\lambda _{s',r',k'}}} \right)}
\end{array}} \right.}  }
\end{equation}
By putting \eqref{MQWLD1} in \eqref{QWCLD} for $T  \to  \infty  $ 
\begin{equation}\label{MQWLD3}
\small{
\begin{array}{l}
\pi \left( v  \right) = \sum\limits_{s,s' = 0}^1 {\sum\limits_{r,r' = 0}^1 {\sum\limits_{\scriptstyle \,\,\,\,\,\,\,\,\, k,k' = 0\hfill\atop
\scriptstyle{\lambda _{s,r,k}} = {\lambda _{s',r',k'}}\hfill}^{N - 1} {{C_{s,r,k}}C_{s',r',k'}^*} } } \\
\,\,\,\,\,\,\,\,\,\,\,\, \times \sum\limits_{b = 0}^1 {\sum\limits_{\rho  = 0}^1 {\left\langle {{b,\rho ,v}}
 \mathrel{\left | {\vphantom {{b,\rho ,v} {{\lambda _{s,r,k}}}}}
 \right. \kern-\nulldelimiterspace}
 {{{\lambda _{s,r,k}}}} \right\rangle } } \left\langle {{{\lambda _{s',r',k'}}}}
 \mathrel{\left | {\vphantom {{{\lambda _{s',r',k'}}} {b,\rho ,v}}}
 \right. \kern-\nulldelimiterspace}
 {{b,\rho ,v}} \right\rangle 
\end{array}
}
\end{equation}
Eigenstates of $\left| {{\lambda _{s,r,k}}} \right\rangle $ are in the k-space so they are in the form of $ \left| {{\chi _{s,r,k}}} \right\rangle \left| {{\kappa _k}} \right\rangle $, so
\begin{widetext}
\small{
\begin{equation} \label{MQWLDNonUniform1}
\begin{array}{l}
\pi \left( v \right) = \sum\limits_{b,s,s' = 0}^1 {\sum\limits_{\rho ,r,r' = 0}^1 {\sum\limits_{\scriptstyle \,\,\,\,\,\,\,\,\,\,k,k' = 0\hfill\atop
\scriptstyle{\lambda _{s,r,k}} = {\lambda _{s',r',k'}}\hfill}^{N - 1} {{C_{s,r,k}}C_{s',r',k'}^*} } } \left\langle {{b,\rho }} \mathrel{\left | {\vphantom {{b,\rho } {{\chi _{s',r',k'}}}}} \right. \kern-\nulldelimiterspace} {{{\chi _{s,r,k}}}} \right\rangle \left\langle {{{\chi _{s',r',k'}}}} \mathrel{\left | {\vphantom {{{\chi _{s,r,k}}} {b,\rho }}} \right. \kern-\nulldelimiterspace} {{b,\rho }} \right\rangle \left\langle {v} \mathrel{\left | {\vphantom {v {{\kappa _k}}}} \right. \kern-\nulldelimiterspace} {{{\kappa _k}}} \right\rangle \left\langle {{{\kappa _{k'}}}} \mathrel{\left | {\vphantom {{{\kappa _{k'}}} v}} \right. \kern-\nulldelimiterspace} {v} \right\rangle 
\\
\\
\,\,\,\,\,\,\,\,\,\,\,\,\, =  \sum\limits_{s,s' = 0}^1 {\sum\limits_{r,r' = 0}^1 {\sum\limits_{\scriptstyle \,\,\,\,\,\,\,\,\,\,k,k' = 0\hfill\atop
\scriptstyle{\lambda _{s,r,k}} = {\lambda _{s',r',k'}}\hfill}^{N - 1} {{C_{s,r,k}}C_{s',r',k'}^*} } }  \left\langle {{{\chi _{s',r',k'}}}} \mathrel{\left | {\vphantom {{{\chi _{s',r',k'}}} {{\chi _{s,r,k}}}}}
 \right. \kern-\nulldelimiterspace} {{{\chi _{s,r,k}}}} \right\rangle \left\langle {{{\kappa _{k'}}}} \mathrel{\left | {\vphantom {{{\kappa _{k'}}} v}} \right. \kern-\nulldelimiterspace} {v} \right\rangle \left\langle {v} \mathrel{\left | {\vphantom {v {{\kappa _k}}}} \right. \kern-\nulldelimiterspace} {{{\kappa _k}}} \right\rangle
\end{array}
\end{equation}
}
\end{widetext}
All the eigenvalues $\lambda_{r,s,k}$ are in the form of $e^{i \phi}$ which can be represented by a point on the unit circle in complex plane. So the degeneracy condition $\lambda_{r,s,k} = \lambda_{r',s',k'}$ in \eqref{MQWLDNonUniform1} reduces to finding two different point on the same position in complex plane. From \eqref{EigVal} it is clear that $\lambda_{0,1,k}$ and $\lambda_{1,1,k}$ belong to quadrant $\mathrm{I}$ and $\mathrm{IV}$ (right zone) and $\lambda_{0,0,k}$ and $\lambda_{1,0,k}$ to quadrant $\mathrm{II}$ and $\mathrm{III}$ (left zone). Since only points from the same zone can occupy same position, therefore the degeneracy conditions reduces to:
\begin{equation}\label{MQWDeg1}
\begin{array}{l}
{\lambda _{0,1,k}} = {\lambda _{0,1,k'}}\\
{\lambda _{1,1,k}} = {\lambda _{1,1,k'}}\\
{\lambda _{0,1,k}} = {\lambda _{1,1,k'}}
\end{array}
\end{equation}
for right zone and

\begin{equation}\label{MQWDeg2}
\begin{array}{l}
{\lambda _{0,0,k}} = {\lambda _{0,0,k'}}\\
{\lambda _{1,0,k}} = {\lambda _{1,0,k'}}\\
{\lambda _{0,0,k}} = {\lambda _{1,0,k'}}
\end{array}
\end{equation}
for left zone. We know $\left\langle {{{\lambda _{r,s,k}}}}  \mathrel{\left | {\vphantom {{{\lambda _{r,s,k}}} {{\lambda _{r',s',k'}}}}} \right. \kern-\nulldelimiterspace} {{{\lambda _{r',s',k'}}}} \right\rangle  = {\delta _{rr'}}{\delta _{rr'}}$, so terms with $r \neq r'$ and $s \neq s'$ do not contribute in summation \eqref{MQWLDNonUniform1}, therefore corresponding degeneracy conditions i.e. ${\lambda _{0,0,k}} = {\lambda _{1,0,k'}}$ and ${\lambda _{0,1,k}} = {\lambda _{1,1,k'}}$ are not necessary to be considered. So by some calculation the whole degeneracy cases can be written as
\begin{equation}\label{MQWDeg3}
\begin{array}{l}
k = k'
\\
\\
k = \frac{n\,N}{2} - k '  \pm \alpha \,\,\,\,\,\,\, n=1,3,...
\end{array}
\end{equation}
Note that we use $\alpha = \frac{N \theta}{2 \pi}$ (called M\"{o}bius factor) as number of complete rotations per cycle.

The interesting result we would like to emphasize here is that degeneracy condition can be controlled by parameter $\alpha$. So for $\alpha \neq \frac{m}{2} \, , \, m=1,2,...$ the term $k = \frac{n\,N}{2} - k '  \pm \alpha $ never satisfies and we don't have degeneracy. So all eigenvalues are distinct. Therefore as discussed in \cite{Aharonov2001} we have a uniform distribution $\pi ( v )= \frac{1}{N}$.

Furthermore, the parameter $\alpha$ can be used to control the rate of convergence to the limiting distribution which defines \textit{mixing time}. 
	Mixing time $M_\epsilon$, measures the number of time steps required for the
	average distribution to be $\epsilon$-close to the limiting distribution \cite{Aharonov2001},
\begin{equation}
{M_\epsilon } = \min \left\{ {T \mid {\forall t \ge T:\left\| {\pi \left( v \right) - {{\bar p}_t}\left( v \right)} \right\| \le \epsilon } } \right\},
\end{equation}
which is one of important differences of QW with CRW. In particular, Aharonov \textit{et al} have shown that odd-sided N-cycles (which has a uniform LD) converges to the limiting distribution in time $O\left(n \log_2 n\right)$, almost quadratically faster than the classical walk. They have proved that, for general quantum walk specified by the unitary matrix $U$, and any initial state $\left|\beta_0\right\rangle = \sum_i{a_i\left|\phi_i\right\rangle}$,
\begin{equation}\label{BoundLD}
\left\| {\pi \left( v \right) - {{\bar p}_t}\left( v \right)} \right\| \le 2\sum\limits_{i,j,{\lambda _i} \ne {\lambda _j}} {{{\left| {{a_i}} \right|}^2}\frac{1}{{t\left| {{\lambda _i} - {\lambda _j}} \right|}}}. 
\end{equation} 
Where $\left|\phi_i\right\rangle$ and $\lambda_i$ are the eigenvectors and corresponding eigenvalues of $U$, respectively. As we see, the distances $\left| {{\lambda _i} - {\lambda _j}} \right|$ are of crucial importance, and they need to be large for the convergence time to be small.
An interesting point in our model is that, the distances $\left| {{\lambda _i} - {\lambda _j}} \right|$ can be controlled by $\theta$ or equivalently $\alpha$ (see \eqref{EigVal}). This means that we are always able to tune $\alpha$ to have uniform LD with optimized mixing time.
	 
In the case  $\alpha = \frac{m}{2} \, , \, m = 0,1,2,...$ the term $k = \frac{n\,N}{2} - k '  \pm \alpha $ can be satisfied in two different situations and degeneracy happens. For odd $N$, $\alpha$ must be half-integer and for even $N$, it must be integer. These two situations lead to non-uniform limiting distributions for which we derive an explicit formula in the next section.

\section{\label{sec:level5}MQW Non-Uniform Limiting Distribution}

Applying degeneracy conditions \eqref{MQWDeg3} into \eqref{MQWLDNonUniform1} and using \eqref{MQWVKappa} leads to
\begin{widetext}
\begin{equation}\label{MQWLD7}
\begin{array}{l}
\pi \left( v \right) =\frac{1}{N} + \frac{1}{N} Re \left( \sum\limits_{\begin{array}{*{20}{c}} {k = 0}\\ {k - \frac{\alpha }{2} \ne \frac{N}{4},\frac{{3N}}{4}} \end{array}}^{N - 1} {{C_{0,0,k}}C_{0,0,\frac{N}{2} + \alpha  - k}^*} \left\langle {{{\chi _{0,0,\frac{N}{2} + \alpha  -k}}}} \mathrel{\left | {\vphantom {{{\chi _{0,0,\frac{N}{2} + \alpha  -k}}} {{\chi _{0,0, k}}}}} \right. \kern-\nulldelimiterspace} {{{\chi _{0,0, k}}}} \right\rangle {e^{ \frac{{2i\pi v}}{N}\left( {2k - \alpha } \right)}}{\left( { - 1} \right)^v} \right)
 \\
 \\
 \,\,\,\,\,\,\, \,\,\,\,\,\,\,\,\,\, \,\,\,\,\,\,\,\,\,\,\,\,\, + Re \left(  \sum\limits_{\begin{array}{*{20}{c}} {k = 0}\\ {k - \frac{\alpha }{2} \ne \frac{N}{4},\frac{{3N}}{4}} \end{array}}^{N - 1} {{C_{0,1,k}}C_{0,1,\frac{N}{2} + \alpha  - k}^*} \left\langle {{{\chi _{0,1,\frac{N}{2} + \alpha  - k}}}} \mathrel{\left | {\vphantom {{{\chi _{0,1,\frac{N}{2} + \alpha  - k}}} {{\chi _{0,1, k}}}}} \right. \kern-\nulldelimiterspace} {{{\chi _{0,1,k}}}} \right\rangle {e^{ \frac{{2i\pi v}}{N}\left( {2k - \alpha } \right)}}{\left( { - 1} \right)^v} \right)
 \\
 \\
 \,\,\,\,\,\,\, \,\,\,\,\,\,\,\,\,\, \,\,\,\,\,\,\,\,\,\,\,\,\, + Re \left( \sum\limits_{\begin{array}{*{20}{c}} {k = 0}\\ {k + \frac{\alpha }{2} \ne \frac{N}{4},\frac{{3N}}{4}} \end{array}}^{N - 1} {{C_{1,0,k}}C_{1,0,\frac{N}{2} - \alpha  - k}^*} \left\langle {{{\chi _{1,0,\frac{N}{2} - \alpha  - k}}}} \mathrel{\left | {\vphantom {{{\chi _{1,0,\frac{N}{2} - \alpha  - k}}} {{\chi _{1,0, k}}}}} \right. \kern-\nulldelimiterspace} {{{\chi _{1,0, k}}}} \right\rangle {e^{ \frac{{2i\pi v}}{N}\left( {2k + \alpha } \right)}}{\left( { - 1} \right)^v} \right)
 \\
 \\
 \,\,\,\,\,\,\, \,\,\,\,\,\,\,\,\,\, \,\,\,\,\,\,\,\,\,\,\,\,\, + Re \left( \sum\limits_{\begin{array}{*{20}{c}} {k = 0}\\ {k + \frac{\alpha }{2} \ne \frac{N}{4},\frac{{3N}}{4}} \end{array}}^{N - 1} {{C_{1,1,k}}C_{1,1,\frac{N}{2} - \alpha  - k}^*} \left\langle {{{\chi _{1,1,\frac{N}{2} - \alpha  - k}}}} \mathrel{\left | {\vphantom {{{\chi _{1,1,\frac{N}{2} - \alpha  - k}}} {{\chi _{1,1, k}}}}} \right. \kern-\nulldelimiterspace} {{{\chi _{1,1, k}}}} \right\rangle {e^{ \frac{{2i\pi v}}{N}\left( {2k + \alpha } \right)}}{\left( { - 1} \right)^v} \right)
\end{array}
\end{equation}
\end{widetext}
We should note that inserting $k=k'$ leads to $\frac{1}{N}$ and since swapping $k$ and $k'$ makes complex conjugate terms in summation, we omit summation over $k'$ and use real part instead. For $N$ divisible by 4, the eigenvalues for $k \pm \frac{\alpha}{2} =\frac{N}{4}$ and $\frac{3N}{4}$ are non-degenerate and unique (see \eqref{EigVal}) , so we have omitted the corresponding terms in summation.

By using explicit from of $| \chi_{x,y,k} \rangle$ (see Appendix.\ref{MQWEigStaApp}) calculations of $C_{x,y,k}C_{x,y,\frac{N}{2} \pm \alpha - k}^*$ and $\langle \chi _{x,y,\frac{N}{2} \pm \alpha - k} | \chi _{x,y,k} \rangle$ are straightforward. We should note that $C_{x,y,k}= \langle \chi_{x,y,k} | \psi_{0} \rangle$ depends on initial state. For example if we use $ | \psi_{0} \rangle  = \tiny{ \left( {\begin{array}{*{20}{c}}
1\\
0\\
0\\
0
\end{array}}\right)}$ as the initial state, the compact form of limiting distribution \eqref{MQWLD7} can be written as
\begin{widetext}
\begin{equation}\label{MQWLD11}
\begin{array}{l}
\small{
\pi \left( v \right) = \frac{1}{N} + \frac{{{{\left( { - 1} \right)}^v}}}{{4{N^2}}}\left( {\sum\limits_{\scriptstyle k = 0\hfill\atop
		\scriptstyle k - \frac{\alpha }{2} \ne \frac{N}{4},\frac{{3N}}{4}\hfill}^{N - 1} {\frac{{\cos \left( {\frac{{2\pi v}}{N}\left( {2k - \alpha } \right)} \right) - \cos \left( {\frac{{2\pi \left( {v + 1} \right)}}{N}\left( {2k - \alpha } \right)} \right)}}{{1 + {{\cos }^2}\left( {\frac{{2\pi k}}{N} - \frac{{\pi \alpha }}{N}} \right)}}}  + \sum\limits_{\scriptstyle k = 0\hfill\atop
\scriptstyle k + \frac{\alpha }{2} \ne \frac{N}{4},\frac{{3N}}{4}\hfill}^{N - 1} {\frac{{\cos \left( {\frac{{2\pi v}}{N}\left( {2k + \alpha } \right)} \right) - \cos \left( {\frac{{2\pi \left( {v + 1} \right)}}{N}\left( {2k + \alpha } \right)} \right)}}{{1 + {{\cos }^2}\left( {\frac{{2\pi k}}{N} + \frac{{\pi \alpha }}{N}} \right)}}} } \right)
}
\end{array}
\end{equation}
\end{widetext}

\begin{figure}
\centering
\includegraphics[width=90mm]{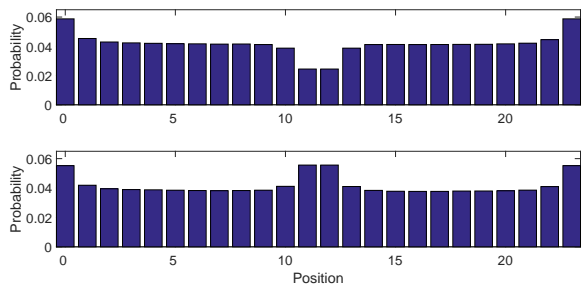} \hfill
\caption{\label{fig24NMQWCLDCombo} LD for Hadamard MQW with 24 nodes and initial state $| \psi _{0} \rangle = | 0 \rangle | 0 \rangle | 0 \rangle $ Up: $\alpha = 1$, Down: $\alpha = 2$ }
\end{figure}

\begin{figure}
\centering
\includegraphics[width=90mm]{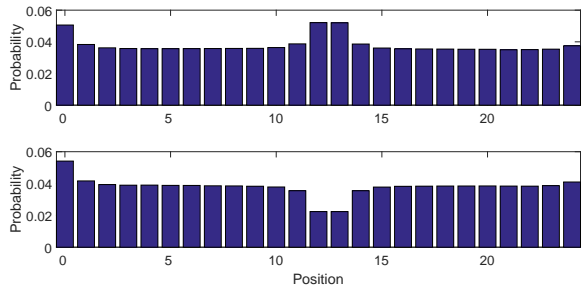} \hfill
\caption{\label{fig26NMQWCLDCombo} LD for Hadamard MQW with 26 nodes and initial state $| \psi _{0} \rangle = | 0 \rangle | 0 \rangle | 0 \rangle $ Up: $\alpha = 1$, Down: $\alpha = 2$ }
\end{figure}

As we expected, for $\alpha = 0$ \eqref{MQWLD11} reduces to \eqref{EvNQWCLD} for QWC. As Bednarska et al. \cite{Bednarska2003} has shown and \eqref{EvNQWCLD} represents, for QWC with even $N$ divisible by 4, we have two hills at $p_{0}$ and $p_{0} + \frac{N}{2}$ Fig. \ref{fig24NQWCLD}. But factor $\alpha$ in \eqref{MQWLD11} can change the scenario. For example for an odd $N$ and $\alpha = \frac{1}{2}$, we have single hill at $p_{0}$ and for an even $N$ and $\alpha = 1,2,... $, depending on $\alpha$ and division of $N$ by 4, we have different forms of limiting distributions Fig.\ref{fig24NMQWCLDCombo}, Fig.\ref{fig26NMQWCLDCombo}.


\section{\label{sec:level6}Conclusion}
We introduced a new type of Quantum Walk on Cycle, namely M\"{o}bius Quantum Walk, in which we added an extra rotation space and defined $\alpha$ which can parameterize number of complete rotations per cycle. We showed that by parameter $\alpha$ we are always able to break degeneracy of evolution operator which leads to a uniform limiting distribution. In other words, it is always possible to reach uniform limiting distribution despite of number of nodes, which was impossible for even number of nodes in QWC.

We also show that the rate of convergence to the limiting distribution can be controlled by M\"{o}bius factor $\alpha$ which it  means that we are always able to tune $\alpha$ to have uniform LD with optimized mixing time.
 Our analysis shows that only for specific values of $\alpha$ limiting distribution is non-uniform for which the explicit from has been driven, showing that we have two hills at $p_{0}$ and $p_{0} + \frac{N}{2}$ when both of conditions $N \, mod \, 4 \, , \, \alpha=2m$ to be satisfied or none of them to be satisfied. We have a hill at $p_{0}$ and a valley at $p_{0} + \frac{N}{2}$ when only one of these conditions to be true.

\appendix
\section{} \label{MQWEigStaApp}
By solving eigenproblem for the $M$ introduced in \eqref{MQWUK}, we will have
\begin{equation} \label{MQWEigStaForm}
\begin{array}{l}
{\chi _{0,0,k}} = \frac{1}{N_{0}} \left[ {\begin{array}{*{20}{c}}
	{{a_0}}\\
	{ - {a_0}}\\
	{{b_0}}\\
	{ - {b_0}}
	\end{array}} \right],{\chi _{0,1,k}} = \frac{1}{N_{1}} \left[ {\begin{array}{*{20}{c}}
	{{a_1}}\\
	{ - {a_1}}\\
	{{b_1}}\\
	{ - {b_1}}
	\end{array}} \right]\\
{\chi _{1,0,k}} =\frac{1}{N_{2}} \left[ {\begin{array}{*{20}{c}}
	{{a_2}}\\
	{ {a_2}}\\
	{{b_2}}\\
	{ {b_2}}
	\end{array}} \right] , \,\,\,\,\, {\chi _{1,1,k}} = \frac{1}{N_{3}} \left[ {\begin{array}{*{20}{c}}
	{{a_3}}\\
	{{a_3}}\\
	{{b_3}}\\
	{{b_3}}
	\end{array}} \right],
\end{array}
\end{equation}
where $N_{i}$ is normalization factor and
	\begin{equation}\label{MQWEigSta}
	\begin{array}{l}
	a_{0}=\sqrt {2}\cos \left( \,\omega_{-} \right) - \frac{1}{\sqrt {2}}\sqrt {1+  \cos ^{2} \left( {\omega_{-}} \right) }+\frac{i}{\sqrt {2}} \sin \left(\omega_{-} \right)
	\\
	a_{1}=- \sqrt {2}\cos \left( \,\omega_{-} \right) - \frac{1}{\sqrt {2}}\sqrt {1+  \cos ^{2} \left( {\omega_{-}} \right)}-\frac{i}{\sqrt {2}} \sin \left(\omega_{-} \right)
	\\
	a_{2}=\sqrt {2}\cos \left( \omega_{+} \right) +\frac{1}{\sqrt {2}} \sqrt {1+  \cos ^{2} \left( {\omega_{+}} \right)}- \frac{i}{\sqrt {2}} \sin \left( \omega_{+} \right)
	\\
	a_{3}=\sqrt {2}\cos \left( \omega_{+} \right) -\frac{1}{\sqrt {2}} \sqrt {1+  \cos ^{2} \left( {\omega_{+}} \right)}-\frac{i}{\sqrt {2}}\sin \left( \omega_{+} \right)
	\\
	b_{0}=- \frac{1}{\sqrt {2}} \sqrt {1+  \cos ^{2} \left( {\omega_{-}} \right)}+ \frac{i}{\sqrt {2}} \sin \left(\omega_{-} \right)
	\\
	b_{1}=\frac{1}{\sqrt {2}}\sqrt {1+  \cos ^{2} \left( {\omega_{-}} \right)} - \frac{i}{\sqrt {2}} \sin \left(\omega_{-} \right)
	\\
	b_{2}=\frac{1}{\sqrt {2}} \sqrt {1+  \cos ^{2} \left( {\omega_{+}} \right)}+\frac{i}{\sqrt {2}}\sin \left( \omega_{+} \right)
	\\
	b_{3}=- \frac{1}{\sqrt {2}} \sqrt {1+  \cos ^{2} \left( {\omega_{+}} \right)}+ \frac{i}{\sqrt {2}} \sin \left( \omega_{+} \right) ,
	\end{array}
	\end{equation}
with $\omega_{\pm }= \frac{\pi \alpha}{N} \pm \frac{2 \pi k}{N}$.

\bibliography{apssamp}

\end{document}